\documentclass[aps,twocolumn,amsmath,amsfonts,nofootinbib,preprintnumbers,superscriptaddress,eqsecnum,secnumarabic]{revtex4-1}
\usepackage{graphicx,slashed,hyperref}
\usepackage[utf8]{inputenc}
\hypersetup{
   bookmarks=true,         
   unicode=true,          
   pdftoolbar=true,        
   pdfmenubar=true,        
   pdffitwindow=false,     
   pdfstartview={FitH},    
   pdftitle={My title},    
   pdfauthor={Author},     
   pdfsubject={Subject},   
   pdfcreator={Creator},   
   pdfproducer={Producer}, 
   pdfkeywords={keyword1} {key2} {key3}, 
   pdfnewwindow=true,      
   colorlinks=true,       
   linkcolor=blue,          
   citecolor=magenta,        
   filecolor=magenta,      
   urlcolor=cyan           
}
\usepackage{amsmath}
\usepackage{amssymb}

\newcommand{\be}{\begin{equation}}
\newcommand{\ee}{\end{equation}}
\def\bsp#1\esp{\begin{split}#1\end{split}}

\begin{document}
\title{
Direct Detection is testing Freeze-in}

\author{Thomas~Hambye}
\email{thambye@ulb.ac.be}
\affiliation{Service de Physique Th\'eorique,
Universit\'e Libre de Bruxelles\\
Bld du Triomphe CP225, 1050 Brussels, Belgium}

\author{Michel~H.G.~Tytgat}
\email{mtytgat@ulb.ac.be}
\affiliation{Service de Physique Th\'eorique,
Universit\'e Libre de Bruxelles\\
Bld du Triomphe CP225, 1050 Brussels, Belgium}

\author{J\'{e}r\^{o}me~Vandecasteele}
\email{jvdecast@ulb.ac.be}
\affiliation{Service de Physique Th\'eorique,
Universit\'e Libre de Bruxelles\\
Bld du Triomphe CP225, 1050 Brussels, Belgium}

\author{Laurent Vanderheyden}
\email{lavdheyd@ulb.ac.be}
\affiliation{Service de Physique Th\'eorique,
Universit\'e Libre de Bruxelles\\
Bld du Triomphe CP225, 1050 Brussels, Belgium}

\date{\today}

\preprint{ULB-TH/18-10}
\begin{abstract}
Dark matter (DM) may belong to a hidden sector (HS) that is only feebly interacting with the Standard Model (SM) and  may have never been in thermal equilibrium in the Early Universe. In this case, the observed abundance of dark matter particles could have built up through a process known as Freeze-in. We show that, for the first time, direct detection experiments are testing this DM production mechanism. 
This applies to scenarios where SM and HS communicate through a light mediator of mass less than a few MeV.  Through the exchange of such light mediator, the very same feebly interacting massive particles can have self-interactions that are in the range required to address the small scale structure issues of collisionless cold DM. 
\end{abstract}

\maketitle

\section{Introduction}

There is overwhelming evidence for dark matter (DM) in the Universe, but its precise nature still eludes us. 
A key question is to explain the observed abundance of DM and several mechanisms have been proposed. A particularly attractive possibility is that DM  is a weakly interacting massive particle (WIMP) that was in thermal equilibrium in the Early Universe, 
down to a temperature at which its number density became so Boltzmann suppressed that it chemically 
decoupled from the thermal bath, 
a process known as Freeze-out (FO). 
The observed relic density is given by the population that was left at the time when DM freezed-out, leading to a DM abundance 
inversely proportional to the cross section responsible for chemical equilibrium
\begin{equation}
Y_{\rm DM} \equiv \frac{n_{\text{DM}}}{s}  \propto {1\over  \sigma}
\end{equation}
where $n_{\rm DM}$ is the DM number density and $s$ the entropy density of the Universe. 
This cross section may be the DM annihilation cross section, or it may be effective, the relevant processes being co-annihilation of the DM with other exotic particles \cite{Griest:1990kh}. Regardless, 
agreement with cosmological observations requires that $ \sigma v \approx 3\times 10^{-26}$ cm$^3$/s, equivalent to a picobarn, characteristic of weak interactions. 
 
This has triggered a vast experimental program that aims at looking for a  WIMP. In particular, 
direct detection experiments search for the recoil of nuclei from collisions with WIMPs from the local halo of DM \cite{Goodman:1984dc,Ahlen:1987mn}. Recently, the XENON1T collaboration has released its latest data \cite{Aprile:2018dbl}. Their strongest constraints on DM-nucleon {spin-independent} (SI) cross section, following a 1 tonne$\times$year exposure, have reached $\sigma_{\rm SI} \approx 7\times 10^{-47}\mathrm{cm}^2$ for a $30\,\mathrm{GeV}$DM particle mass, setting a new landmark in the quest for a WIMP.  In this short note, we put forward the fact that, for the first {time,  direct detection experiments, in particular XENON1T, have} also reached the sensitivity required to test another mechanism for the DM abundance: Freeze-in (FI).

The FI mechanism rests on the possibility that a DM particle may be only feebly interacting with the known SM, hence a feebly interacting massive particle (FIMP). Unlike a WIMP, a FIMP may have never been in thermal equilibrium in the Early Universe \cite{Hall:2009bx,Dodelson:1993je,McDonald:2001vt}. Instead, its abundance  could have slowly built up through SM particles collisions, for example SM + SM $\rightarrow$ DM + DM pair creation. In this case
\begin{equation}
Y_{\rm DM} \propto  \sigma
\end{equation} 
As it is well known, the abundance of DM is typically a very small number,  $Y_{\rm DM}~\approx 4\times 10^{-10}({\rm GeV}/m_{\rm DM})$. 
In the case of FO, this is explained by Boltzmann suppression of the DM abundance at the time of thermal decoupling. In the case of FI mechanism, this is  simply provided by a very small cross section for DM production. Indeed, focusing on the case of renormalizable interactions, the required coupling must be tiny, typically of order $10^{-10}$. 
As simple and natural as the freeze-in scenario may be, given the smallness of such coupling, one may wonder if we will ever be able to test it experimentally? 
We emphasize here the fact that FI is already being tested by XENON1T for the important class of FIMPs for which the SM and DM are coupled through a light mediator particle. 
The exchange of such a light mediator between DM particles can also induce large self-interaction, a possibility  that is of much interest because it may alleviate the possible shortcomings of collisionless dark matter on small scales (core/cusp \cite{Spergel:1999mh} and too-big-to-fail \cite{BoylanKolchin:2011de} problems, see \cite{Tulin:2017ara} for a review). 
We show that indeed several FIMP candidates that can meet these additional issues are also being tested by XENON1T.

\section{Framework}
\label{sec:framework}

We consider the slow, out-of-equilibrium production of DM  from the thermal bath of SM particles during the radiation dominated era.
Typical processes include DM pair production from one (SM $\rightarrow$ DM + DM or $1\rightarrow 2$) or two SM particles (SM + SM $\rightarrow$ DM + DM or $2\rightarrow 2$).
The general FI scenario we refer to here is based on the assumption that these processes are mediated by particles lighter than the SM and DM particles involved (see \cite{Dodelson:1993je,Blennow:2013jba} for non-thermal production through heavy mediators). In this case, the DM is essentially produced at a temperature $T$ approximately equal to the mass of the heaviest particle involved, $T_{\rm FI}\equiv max[m_{\rm DM},m_{\rm SM}]$ \cite{Hall:2009bx,McDonald:2001vt,Frigerio:2011in,Chu:2011be}. Hence, such FI is an infrared dominated production mechanism, which rests only on dynamics around  $T \sim T_{\rm FI}$. The fact that DM is dominantly produced around $T_{\rm FI}$  stems from the $T$ dependence of the rate for DM production (per unit volume), $\gamma_{1\rightarrow 2}$ and/or $\gamma_{2\rightarrow 2}$, relative to the Hubble expansion rate $H$ (times the entropy density).
In particular, at $T \lesssim T_{FI}$ the production rate decreases rapidly because of Boltzmann suppression:
if $m_{\rm DM}>m_{\rm SM}$ this is due to exponential suppression of 
the number of SM particles with  energy large enough to produce DM particles; if instead
$m_{\rm DM}<m_{SM}$, it is the abundance of SM particles that is Boltzmann suppressed.
The number of DM particles produced around $T\sim T_{FI}$ is thus essentially determined by the rate of production at $T_{\rm FI}$ times the age of the Universe at $T_{\rm FI}$, $t_{\rm FI} \approx 1/H(T_{\rm FI})$. For instance, for the case of annihilation,
\begin{equation}
\label{eq:abundance}
Y_{\rm DM} \equiv \frac{n_{\rm DM}}{s} \simeq  \frac{\gamma_{2 \rightarrow 2}}{s H}\Big|_{T_{\rm FI}} = \frac{n_{\rm SM,eq}^{2} \langle \sigma_{2 \rightarrow 2} \, v \rangle}{sH} \Big|_{T_{\rm FI}}
\end{equation} 
where $n_{\rm SM,eq}$ is the abundance of the relevant SM particles at equilibrium. 
Observations require $Y_{\rm DM}~\approx~4\times~10^{-10}({\rm GeV}/m_{\rm DM})$, a very small number that is easily achieved if the underlying interactions are feeble. 
For instance, if the cross section in (\ref{eq:abundance}) is quadratic in a coupling parameter, say, $\kappa$, then typically one needs $\kappa = {\cal O}(10^{-10})$ as we will see in an explicit model below. Let us emphasize here that FI and FO are complementary mechanisms, corresponding to opposite corners of the  parameter space of a DM model. Starting from FI, with a tiny effective coupling $\kappa$ driving a $2 \rightarrow 2$  cross section, one can  reach the FO regime by continuously increasing  $\kappa$. The DM abundance first increases with $\kappa$, eventually overshooting the observed relic abundance, and does so as long as the $2 \rightarrow 2$ does not reach equilibrium.  When $\kappa$ is large enough, DM is in thermal equilibrium and one enters the FO regime, in which the relic density decreases with $\kappa$ \cite{Hall:2009bx,Chu:2011be}.
 
 In the sequel, we focus on a class of FI scenarios with a rather light mediator, with mass below ${\cal O}$(MeV). The reason that such FI scenarios can already be tested by current direct detection experiments stems from the fact, already emphasized in \cite{Chu:2011be,Bernal:2015ova}, that DM elastic collisions with nuclei proceed through the t-channel, which for a light enough mediator
results in a boost of the cross section that can largely compensate the smallness of the coupling required for FI.  
 To illustrate this, we consider the particularly simple model of millicharged DM.  In its minimal version, millicharged DM consists of a single massive particle ($\chi$ in the sequel) charged under a new $U(1)'$ gauge symmetry, with fine structure constant $\alpha^\prime=e'^2/4\pi$. Just like the electron in the SM, the stability of $\chi$ is guaranteed by gauge symmetry (see {\em e.g.} \cite{Hambye:2010zb}).  The $\chi$ field naturally couples to the SM through mixing of its gauge field strength tensor with the one of $U(1)_Y$ hypercharge gauge field
 \begin{equation}
 {\cal L} \supset - {\varepsilon\over 2} F_Y^{\mu \nu} F^\prime_{\mu \nu},
 \end{equation}
a mechanism known as kinetic mixing \cite{Holdom:1985ag}. Because of kinetic mixing, the $\chi$ particle (and its antiparticle) couples to the photon with a millicharge
 \begin{equation}
 \label{eq:millicharge}
 \kappa=(e'/e) \epsilon \cos \theta_\epsilon/\sqrt{1-\epsilon^2}
 \end{equation}
where $\tan \theta_\epsilon = \tan{\theta_W}/\sqrt{1-\epsilon^2}$ and $\theta_W$ is the Weinberg angle; $\kappa$ is the 
parameter that controls the coupling between the DM and electrically charged SM particles, hence both FI production cross sections and the direct detection elastic collisions.
This simple model is one instance of a generic scenario in which DM belongs to a hidden sector (HS) that is coupled with the SM through one of its so-called portals \cite{Patt:2006fw}. 
Clearly, the HS may be more complex, containing more than one massive particle (including a full mirror copy of the SM \cite{Kobzarev:1966qya,Foot:2004pa}). Also, the $U(1)^\prime$ may or may not be spontaneously broken. In the latter case, the SM and HS particles interact feebly with each other through exchange of massive gauge bosons, one being the ordinary $Z$ gauge boson, the other one being a new $\gamma'$ state, also called dark photon 
(see {\em e.g.} \cite{Chu:2011be}).

\section{Recasting direct detection limits}
\label{sec:recast}
Direct detection experiments report their results in the plane $m_{\rm DM}-\sigma_{{\rm DM},n}$ where $n$ stands for a nucleon (typically the neutron). Specifically, $\sigma_{{\rm DM}, n}$ refers to a total elastic collision cross section, which is considered to be either  spin-dependent or spin-independent. We focus on the latter because it provides the strongest constraints and it is the one relevant for the FIMP scenario described at the end of section \ref{sec:framework}. 

The XENON1T collaboration has published their so far strongest constraints on DM-nucleon SI cross sections following a 1 tonne$\times$year exposure, pushing it down to $7\times 10^{-47}\, \mathrm{cm}^2$ for a $30\,\mathrm{GeV}$ DM mass. The XENON1T limits have been derived assuming the scattering as being due to a short range contact interaction, mediated by a heavy mediator.
Here we consider the possibility that the interaction between a DM particle and a nucleus N  is  mediated by a light $\gamma'$, so as to lead to a long-range interaction, with  t-channel propagator
\begin{equation}
{1\over t- m_{\gamma^\prime}^2} = \frac{-1}{ 2m_N E_R+ m^2_{\gamma'}}
\end{equation}
where $m_N$ is the nucleus mass (\textit{e.g.} a Xenon nucleus) and $E_R$ is the recoil energy probed in direct detection experiments, which is typically $\gtrsim 5$ keV. 
If 
 \begin{equation}
\label{eq:lightmed}
 m_{\gamma^\prime}~<~\sqrt{2m_N E_R}~\sim{\cal O} {\rm (40MeV)}
 \end{equation}
 the DM-N differential cross section scales as $1/E_R^2$, which can lead to a large enhancement of the collision rate at small $E_R$.
Because of this, the XENON1T results do not directly apply to the case of a light mediator. 
Nevertheless, it is possible to get rather conservative limits by recasting their constraints based on an $E_R$ independent cross section.
To this end, we begin with the differential rate of collisions (the number of events per second per unit of recoil energy)
\begin{align}
\frac{\mathrm{d}R}{\mathrm{d}E_R}={N_T \,n_{\rm DM} \int \frac{\mathrm{d}\sigma}{\mathrm{d}E_R} {v} f_{\mathrm{\oplus}}\left(\vec{v}\right)\mathrm{d}^3v}
\label{eq:DiffRate}
\end{align}
with $N_T$ the number of targets, $n_{\rm DM}$ the local number density of DM and $f_{\oplus}\left(\vec{v}\right)$ its velocity distribution in the Earth frame, which we take to be Maxwellian with r.m.s velocity $\sigma_v=270\text{ km}/\text{s}$ in the Galactic reference frame. In (\ref{eq:DiffRate}),  $d\sigma/d E_R$ is the DM-N differential cross section for a given recoil energy $E_R$. The integration is made on all velocities that are allowed for a given $E_R$, that is to say satisfying 
$
v \geq v_{\text{min}}=\sqrt{{m_N E_R}/{2\mu_{\chi N}^2}}
$
with $\mu_{\chi N}$ the DM-N reduced mass. 
It is convenient to write $d\sigma/dE_R$ as \cite{Fornengo:2011sz}
\begin{align}
\frac{\mathrm{d}\sigma}{\mathrm{d}E_R}=\frac{m_N}{2\mu_{\chi p}^2}\frac{1}{v^2}\sigma_{\chi p}(E_R) Z^2 F^2\left(q\, r_A\right)
\label{eq:dsigmadE}
\end{align}
where $\mu_{\chi p}$ is the DM-proton reduced mass and $Z$ the nucleus atomic number. The $F\left(q r_A\right)$ is the nucleus form factor, which for concreteness we take from \cite{Fornengo:2011sz,PhysRev.104.1466}.  For a SI contact interaction and no isospin violation,  we can simply set $\sigma_{\chi p} \equiv \sigma_{\chi n}  = \sigma_{{\rm DM}, n}$ and replacing $Z$ by the mass number $A$. Instead, in the millicharged scenario and dark photon with mass $m_{\gamma^\prime}$
\begin{align}
\sigma_{\chi p}(E_R)=\frac{16\pi \mu^2_{\chi p}\alpha^2\kappa^2}{\left(2m_NE_R+m_{\gamma'}^2\right)^2}
\label{eq:LM}
\end{align}
where $\alpha$ is the QED fine structure constant. 
\begin{figure}[h]
\includegraphics[width=8cm]{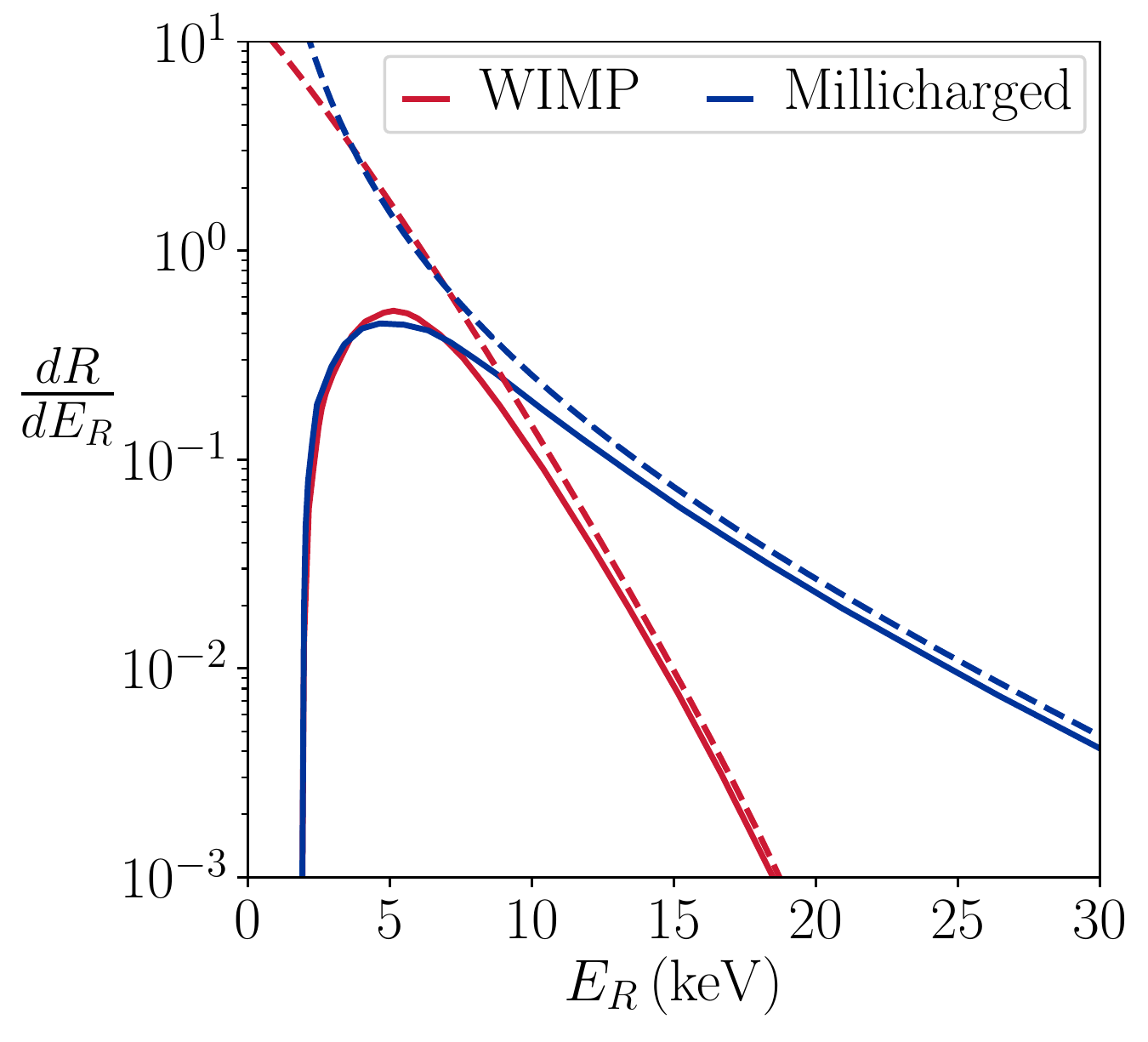}
\caption{Red solid: differential rate for a $\left(m_{\text{DM}},\sigma_{{\rm DM},n}\right)=\left(15 \text{ GeV}, 10^{-46}\text{ cm}^2\right)$. Blue solid: best fit for a DM candidate with long range interactions with $\left(m_\chi,\kappa\right)=\left(70\text{ GeV},3.1\times 10^{-11}\right)$. The error  is $\Delta_{\text{DR}} \approx 25\%$. The dashed curves are for the same candidates, but not taking into account XENON1T efficiency.}
\label{fig:DiffRateFit}
\end{figure}

If (\ref{eq:lightmed}) is satisfied, it is most natural to express the direct detection constraints in the plane $m_{\chi}-\kappa$. To do so, we can exploit two facts. First, while the differential rates through a light and massive mediator have distinct shapes for DM candidates of {\em same} mass, they can be similar for {\em distinct} DM masses, at least within a given recoil energy $E_R$ range. Second, due to detection efficiency, nuclear form factor and velocity distribution, the differential rates fall rapidly at low and high recoil energies and this regardless of the type of interactions, so only a finite range of $E_R$ is relevant for each candidates. These allow to map $(m_{\rm DM},\sigma_{{\rm DM},n})$ onto $(m_\chi,\kappa)$. 
In practice, we take the upper limit on $\sigma_{{\rm DM}, n}$ from XENON1T corresponding to a  given DM candidate, to get a proxy for the observable differential rate, 
\begin{align}
\Big(\frac{\mathrm{d}R}{\mathrm{d}E_R}\Big)_{\text{exp}}=\epsilon\left(E_R\right) \frac{\mathrm{d}R}{\mathrm{d}E_R},
\label{dRdEmeas}
\end{align}
where  $\epsilon(E_R)$ is the detector efficiency from Fig.~1 in \cite{Aprile:2018dbl} and with ${\mathrm{d}R}/{\mathrm{d}E_R}$ given by (\ref{eq:DiffRate}) and (\ref{eq:dsigmadE}) with a constant cross section  $\sigma_{{\rm DM},n}$ and $Z \rightarrow A$.
For this candidate $(m_{\rm DM}, \sigma_{{\rm DM},n})$, we then determine  the couple $(m_{\chi}, \kappa)$  that would have a similar observable differential rate $(dR_{\chi}/d E_R)_{\rm exp}$. Concretely,  we  consider the couple $(m_\chi, \kappa)$  that minimizes the quadratic distance between the two rates
\begin{align}
\Delta^2_{\rm DR} = {1\over R_{\rm exp}^2} \int \mathrm{d}E \,\epsilon\!\left(E\right)^2 \left(\Big(\frac{\mathrm{d}R}{\mathrm{d}E}\Big)-\Big(\frac{\mathrm{d}R_\chi}{\mathrm{d}E}\Big)\right)^2 
\label{eq:Error}
\end{align} 
where $R_{\text{exp}}$ is the total measurable rate from (\ref{dRdEmeas}) and $dR_{\chi}/d E_R$ is the differential rate corresponding to a candidate with a light mediator.\\

This procedure is illustrated by Fig.~\ref{fig:DiffRateFit}. The dashed red line represents the differential rate obtained for a constant cross section ($\sigma_{{\rm DM},n}=10^{-46}\text{ cm}^2$) and  a DM mass ($m_{\rm DM} = 15$ GeV) taken from the XENON1T upper bound. The dashed blue line is the differential rate obtained from the  cross section (\ref{eq:LM}) that minimizes $\Delta_{\text{DR}}$, here corresponding to the candidate $\left(m_\chi,\kappa\right)=\left(70\text{ GeV},3.1\times 10^{-11}\right)$. The solid lines correspond to the observable differential rates, taking into account efficiency. We emphasize that the matching is dominated by low recoil energies $E_R<10\text{ keV}$ corresponding to higher expected event numbers.  In this example, $\Delta_{\rm DR} \approx 25 \%$ while over all the DM mass range we consider,  $\Delta_{\rm DR}$ never exceeds ${\cal O}(30 \%)$. {The same minimization criterion implies that the $70$ GeV $\chi$ particle and the $15$ GeV XENON1T DM candidate have very similar total number of events, in the present case
\begin{align}
\Delta_{\rm TR} = { N_{70}^{\chi} - N_{15}^{\text{DM}}\over N_{15}^{\text{DM}}} \approx 20 \%
\end{align}
where, say, $N_{70}^{\chi}$ is the total number of events (taking into account acceptance) expected for the $\left(m_\chi,\kappa\right)=\left(70\text{ GeV},3.1\times 10^{-11}\right)$ candidate. For reference, we give in Table \ref{table1} such correspondences with XENON1T limits for three distinct $\chi$ masses. We emphasize that the errors on the total rate are positive, meaning that the total number of events is larger for the $\chi$ particle than for the corresponding DM particle with a massive mediator. Reducing the error on the total number of events would thus require decreasing the parameter $\kappa$ (at the expense of the matching between the differential rates). In that sense, we deem our constraints on $\kappa$ to be conservative. {Here we focus on the fact that current data are testing the freeze-in scenario  for DM candidates in the multi-GeV range. In the future, the FIMP scenario could also be tested in the sub-GeV range and with very light mediators, using scattering of the $\chi$ particles on electrons instead of nuclei and the expected yearly modulation of DM collisions to tame the experimental background \cite{Essig:2017kqs} or new technology for detectors \cite{Knapen:2017ekk}. Notice that for the "heavy" DM candidates we consider, and for mediators in the MeV range, the elastic cross-section on electrons is extremely small, $\sigma_{\chi e} \sim 10^{-50}$ cm$^2$, way beyond the reach of current experiments.}}
\begin{table}
\begin{tabular}{c|c||c|c||c|c}
  $m_\chi$ (GeV)& $\kappa\, (10^{-11})$ & $m_{\rm DM}$ (GeV) &$\sigma_{{\rm DM},n}$ (cm$^2$) & $\Delta_{\rm DR}$ & $\Delta_{\rm TR}$   \\
 \hline
  $15$  & $3.0$ & $10$ & $5.6\times 10^{-46}$ &   $16 \%$ &  $15 \%$  \\
$70$   & $3.1$ & $15$ & $1.1\times 10^{-46}$ & $23 \%$ & $22 \%$\\
 $200$  & $5.2$  &  $20$& $5.9\times 10^{-47}$ &  $22 \%$ &  $13 \%$  \\
 $ 500$ & $8.2$ & $22 $& $5.3 \times 10^{-47} $& $26\%$& $3\%$
\end{tabular}
\caption{Upper bounds on the mixing parameter $\kappa$ (2nd column) for three $\chi$ masses (1st column), based on the correspondence with WIMP exclusion limits (3rd and 4th columns). The last two column give respectively the error on the differential rate $\Delta_{\rm DR}$ and total rate $\Delta_{\rm TR}$. 
\label{table1}
}
\end{table}

\section{Freeze-in {\em vs} direct detection}
\label{sec:results}

\begin{figure}[h!]
\includegraphics[width=8cm]{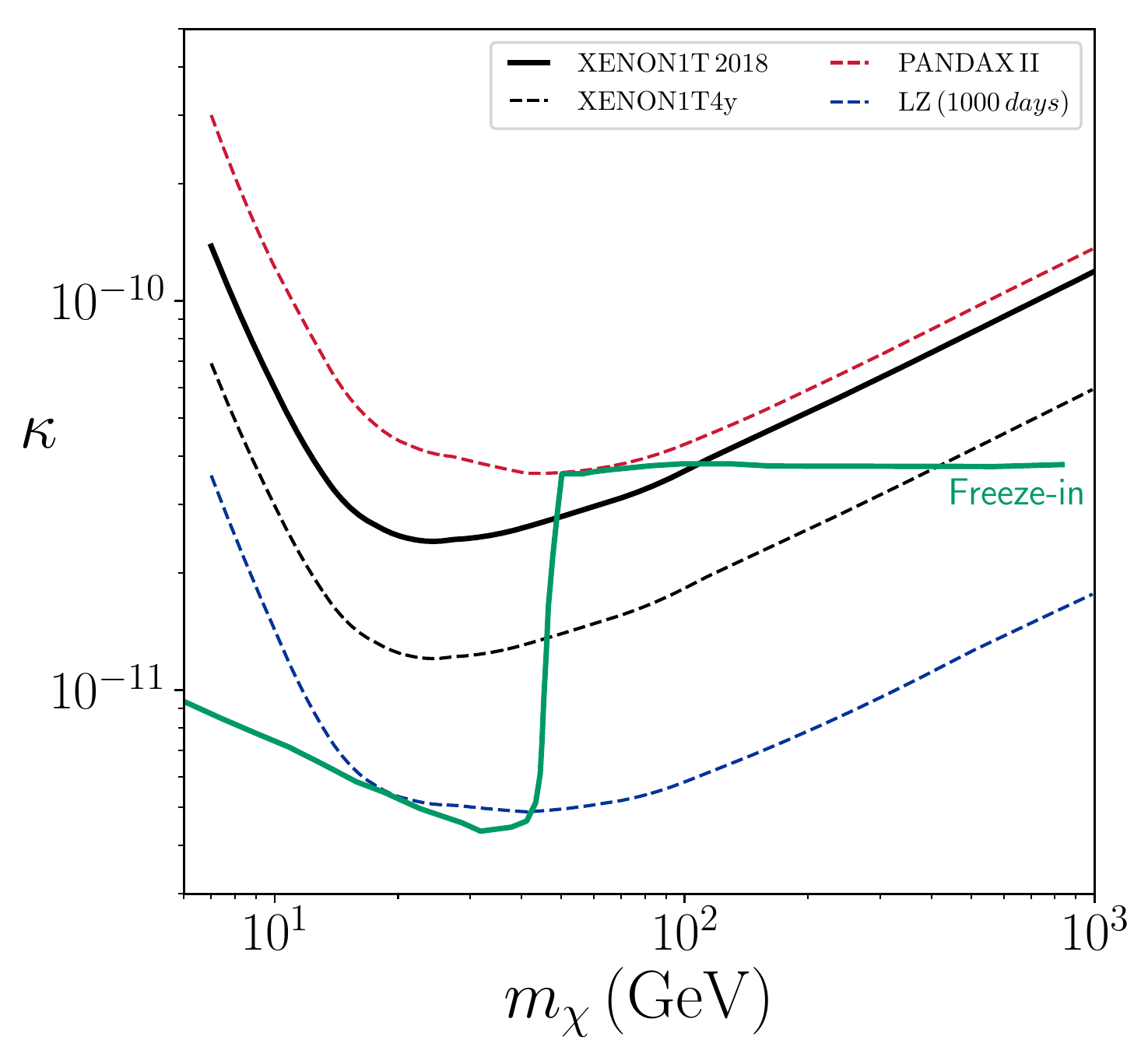}
\caption{Exclusion limits from XENON1T (black), forecast for XENON1T for 4 years (black, dashed), PANDAXII (red, dashed) and  forecast for LZ for 1000 days (blue, dashed). The solid green line corresponds to the  $\kappa$ needed to reproduce the observed relic density through the FI mechanism.}
\label{fig:Constraints_plot}
\end{figure}

Scanning over the XENON1T exclusion limits, the solid black line in Fig.~\ref{fig:Constraints_plot} gives the upper limits on the coupling $\kappa$ as a function of the DM mass $m_\chi$.
We emphasize that these limits only use the XENON1T constraints in the range $1$ GeV $\lesssim m_{\rm DM} \lesssim  50$ GeV (see Table \ref{table1}). This stems from the $\propto 1/E_R^2$ behavior of the cross section, leading to events in the low $E_R$ region, which for a heavy mediator corresponds to relatively lighter DM candidates.
In the same Fig.~\ref{fig:Constraints_plot}, the solid green line gives the $\kappa$
corresponding to the observed DM abundance, along the FI scenario depicted in section \ref{sec:framework} (see \cite{Chu:2011be}). In the millicharged model, FI is set 
by annihilation of SM particles into DM pairs, $f\bar{f}/W^{+}W^{-}\rightarrow {\chi}$ ${\bar \chi}$,  and by $Z$ decay, $Z\rightarrow {\chi}$ ${\bar\chi}$ \cite{Chu:2011be,Essig:2011nj}.  {The dip  at $m_{\chi}\simeq m_{Z}/2$ corresponds to production of $\chi \bar \chi$ from decay of on-shell $Z$ bosons, which is the dominant channel for $10^2\; {\rm MeV} \lesssim  m_{\chi} \leq m_Z/2$. Above $m_Z/2$, production is through both virtual dark photon and Z exchange \cite{Chu:2011be,Essig:2011nj}.}

{Fig.~\ref{fig:Constraints_plot} reveals that  XENON1T is testing for the first time a FI scenario, excluding millicharged FIMP candidates 
within the $m_Z/2 < m_{DM} < 100$~GeV range. }
We also show the limits from the  2017 PANDAXII results \cite{Cui:2017nnn}, following the same procedure we used for recasting XENON1T data.
PANDAXII limits almost reach the FI parameter range. 
Finally, we show the prospects  for XENON1T for 4 years of exposure and for the future LZ experiment \cite{Akerib:2018lyp} (for 1000 days). 
XENON1T should  probe the millicharged FI scenario  for $m_\chi$ from 45~GeV up to $\sim 400$~GeV, whereas LZ could test it all the way from $m_{\rm DM} \sim15$~GeV to $\sim 4$~TeV, {potentially also testing freeze-in from Z decay}.

We have so far neglected the mass of the dark photon, an approximation which is valid as long as $m_{\gamma'}\lesssim  \sqrt{2m_N E_R} \sim 40\text{ MeV}$, taking $E_R$ to be around $5$ keV as typical recoil energy, see Fig.~\ref{fig:DiffRateFit}. Thus for a MeV dark photon, our results still apply. As soon as $m_{\gamma'} \gtrsim 40\text{ MeV}$, the collinear enhancement is lost, which results in less stringent constraints on $\kappa$ and thus no direct detection test of  FI.
\footnote{Nevertheless, in this case, the current direct detection experiments can still probe the so-called re-annihilation regime, a DM production mechanism intermediate between FO and FI \cite{Chu:2011be}.} 
Indeed,  the FI abundance itself is insensitive to the mass of the dark photon, at least provided  $m_{\gamma^\prime} \leq 2 \,m_\chi$, values that lie outside the parameter range we consider here.

\begin{figure}[!h]
\begin{minipage}[t]{1.0\linewidth}
    \centering
    \includegraphics[width=8cm]{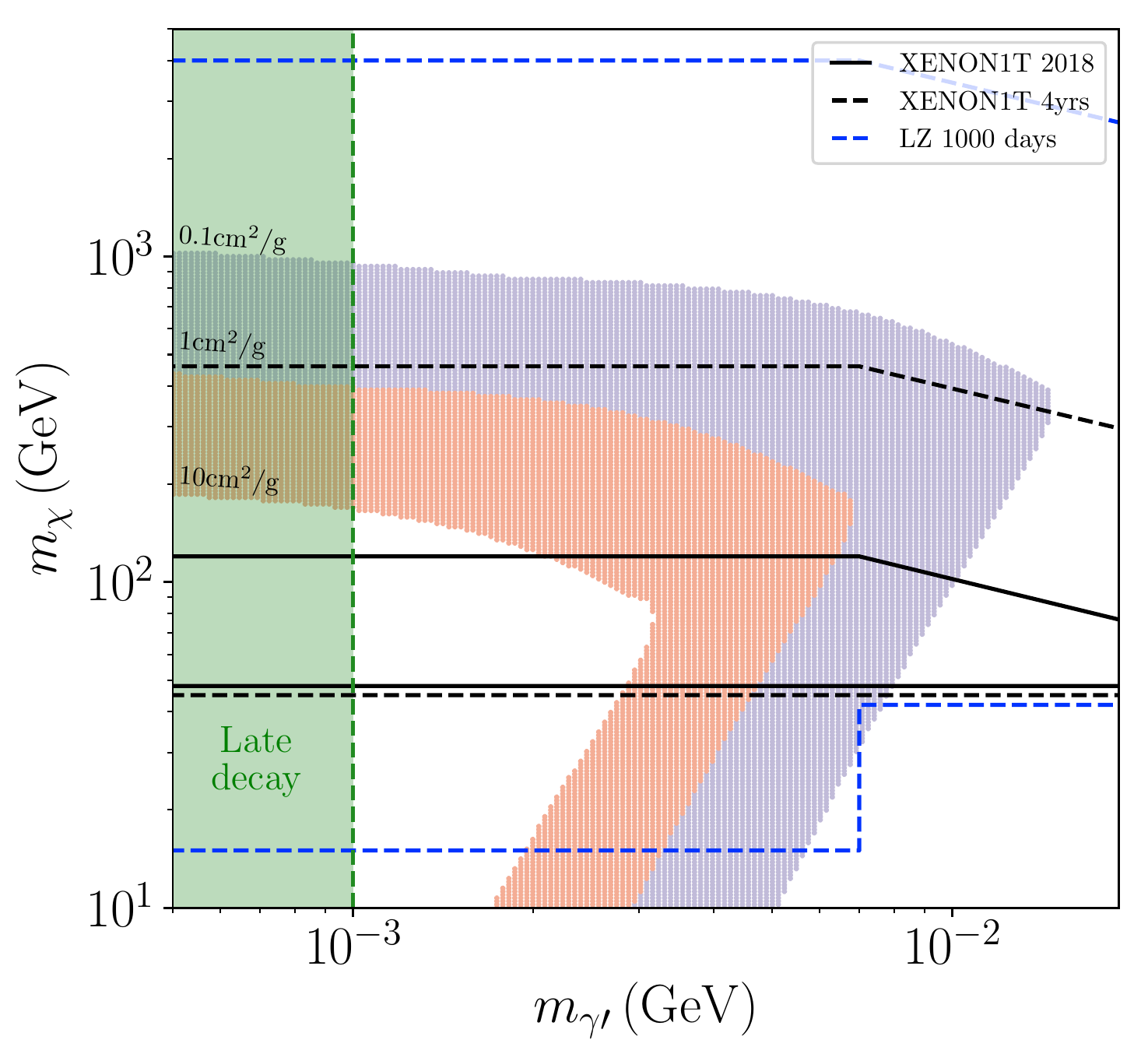}
\end{minipage}
\begin{minipage}[t]{1.0\linewidth} 
    \centering
    \includegraphics[width=8cm]{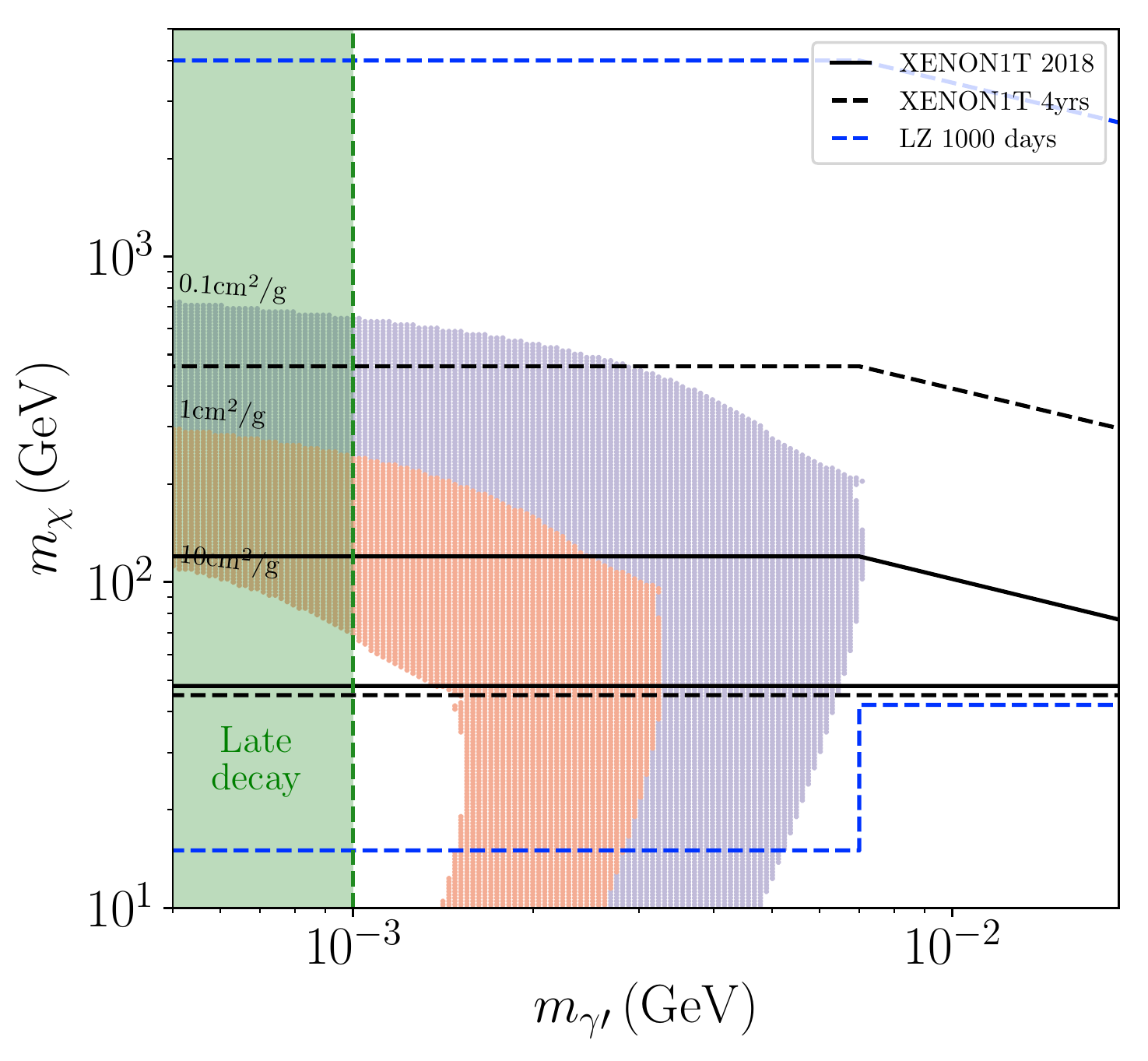}
    \label{f2}
\end{minipage}
\caption{Parameter space for fixed DM coupling and $\alpha^\prime=5\times 10^{-5}$ showing DM candidates with attractive (top) and repulsive (bottom) self-interaction that can alleviate the small-scale structure issues. Red: $1 \text{ cm}^{2}/\text{g}\leq\sigma _T/m_\chi \leq 10 \text{ cm}^{2}/\text{g}$; Blue: $0.1 \text{ cm}^{2}/\text{g}\leq\sigma _T/m_\chi \leq 1 \text{ cm}^{2}/\text{g}$. The lines delimitate the FIMP region probed by the current XENON1T (black solid), XENON1T after 4 years of exposure (black dashed) and the future LZ (blue dashed).}
\label{fig:SIDM}
\end{figure}

\section{Self-interacting dark matter}
\label{sec:bs}

A massless dark photon leads to infinite range forces, both between DM particles and between DM and ordinary matter. Such interactions could be in conflict with observations of the dynamics of galaxies \cite{Feng:2009mn,Ackerman:mha}. Also, a millicharged DM would be deflected by the magnetic field in galaxies  \cite{Chuzhoy:2008zy,McDermott:2010pa,SanchezSalcedo:2010ev} or clusters  \cite{Kadota:2016tqq}.  In particular, a millicharged particle from the DM halo could be repelled by the coherent magnetic field in our Galaxy, which could deplete the local abundance of DM, thus decreasing the reach of direct detection experiments \cite{Chuzhoy:2008zy}. A massive dark photon in the range ${\cal O}$(MeV) circumvents all these {potential} complications  \cite{Chu:2011be}. This may also be a blessing as a $\sim$~MeV mediator can induce a large self-interaction cross section (noted here $\sigma_T$). Provided $0.1 \text{cm}^2/\text{g}<\sigma_T/m_\chi <10\text{cm}^2/\text{g}$, this could alleviate the  small scale structure issues of collisionless cold dark matter \cite{Tulin:2013teo,Tulin:2017ara}. 

Fig.~\ref{fig:SIDM} shows the candidates that satisfy this requirement for $\alpha ' = 5\times 10^{-5}$ and setting the velocity of DM particles to $10$ km/s, relevant for the core/cusp problem in dwarf spheroidal galaxies \cite{Tulin:2017ara}. This value of $\alpha^\prime$ is chosen so that over all the mass we consider, the candidates are in the freeze-in regime (see Ref.\cite{Chu:2011be}). 
The black solid line encloses FIMP candidates that  XENON1T is currently testing and the dashed lines are for XENON1T with hypothetical 4 years of exposure (black) and LZ for $10^3$ days (blue). Here, we show both the attractive (top) and repulsive (bottom) channels. Clearly, there is a large overlap between the regions of $m_{\chi} - m_{\gamma^\prime}$ satisfying the self-interactions constraint, satisfying the relic density constraint through FI, and that can be tested by current and future direct detection experiments. 

We emphasize that self-interacting DM scenarios based on FO and a light mediator are severely constrained by 
CMB data \cite{Bringmann:2016din,Bernal:2015ova}. {There are also cosmological and, potentially, astrophysical constraints that are relevant for the FI scenario with a mediator in the MeV range. As long as $m_{\gamma^\prime} \gtrsim 2 m_e$,  dark photons, which are  produced through freeze-in with $\kappa \sim 10^{-10}$ and thus have an abundance smaller than in thermal equilibrium, decay into $e^+ e^-$ without affecting the predictions from big bang nucleosynthesis (BBN) \cite{Berger:2016vxi}. Below the $e^+ e^-$ threshold however, dark photon can decay into three photons at one loop level, with a rate  that is however excluded by the diffuse X-ray background \cite{Chang:2016ntp,Essig:2013goa}.  This constraint is shown in Fig.~\ref{fig:SIDM} as the greenish region (late decay). Interestingly, the FI scenario with a $\sim $ MeV mediator is potentially constrained by the possibility of emitting dark photons during supernovae collapse, and more specifically SN1987A \cite{Kazanas:2014mca}. The physics of supernovae is however subject to large systematic uncertainties and such constraints are much less robust that those based on cosmological production and decay of dark photon. Moreover, the production of dark photons in supernovae have been revisited in more recent works \cite{Chang:2016ntp,Mahoney:2017jqk,Chang:2018rso}, which claim that thermal and finite density effects, that have been neglected in \cite{Kazanas:2014mca}, may strongly affect their results. Clearly, it would be of interest to try and clarify the status of such constraints, especially given the interplay between the freeze-in mechanism, direct detection searches and self-interacting dark matter. }

\begin{figure}[!h]
\begin{minipage}[t]{1.0\linewidth}
    \includegraphics[width=8cm]{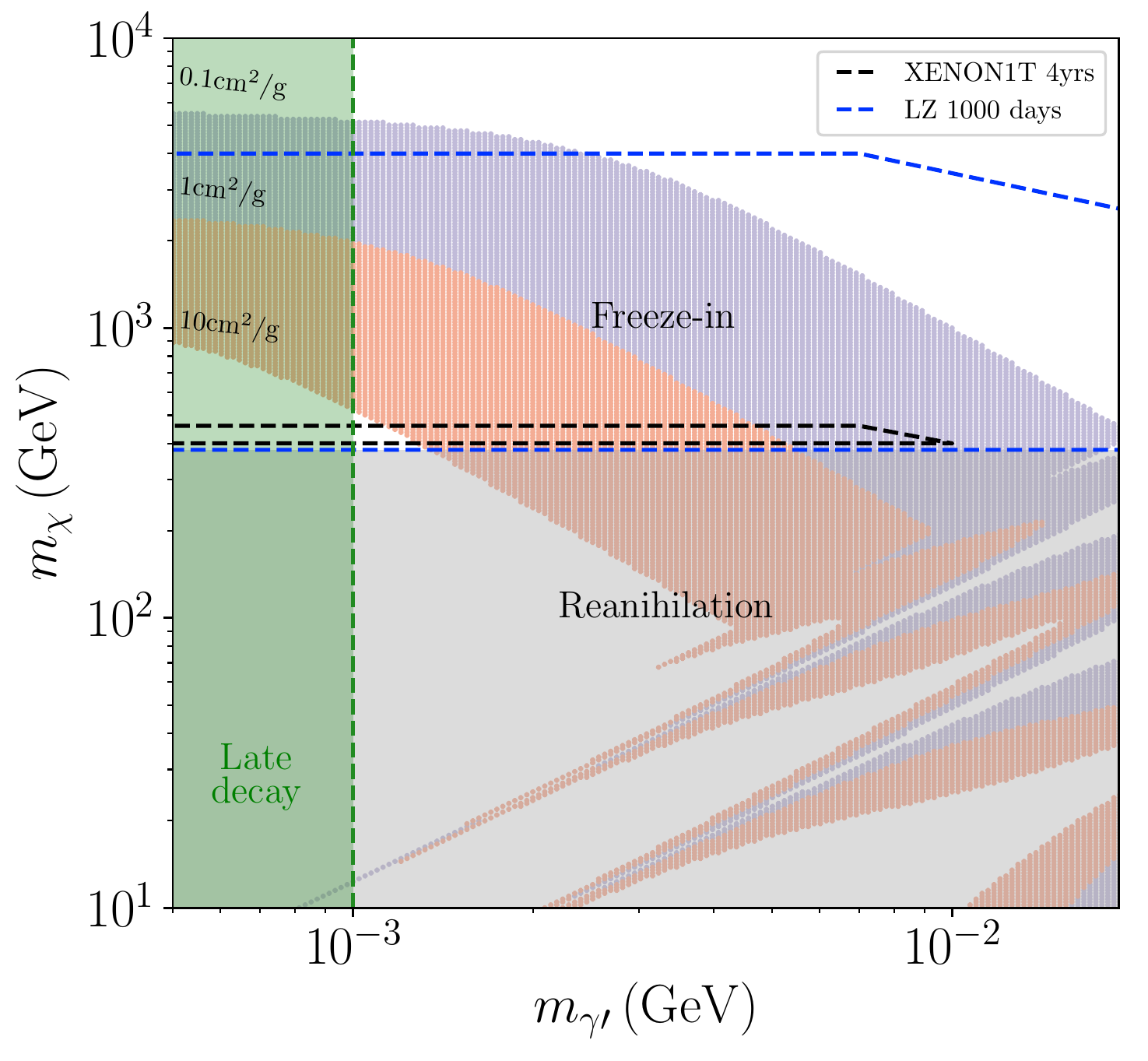}
    \label{f1}
\end{minipage}
\begin{minipage}[t]{1.0\linewidth} 
    \includegraphics[width=8cm]{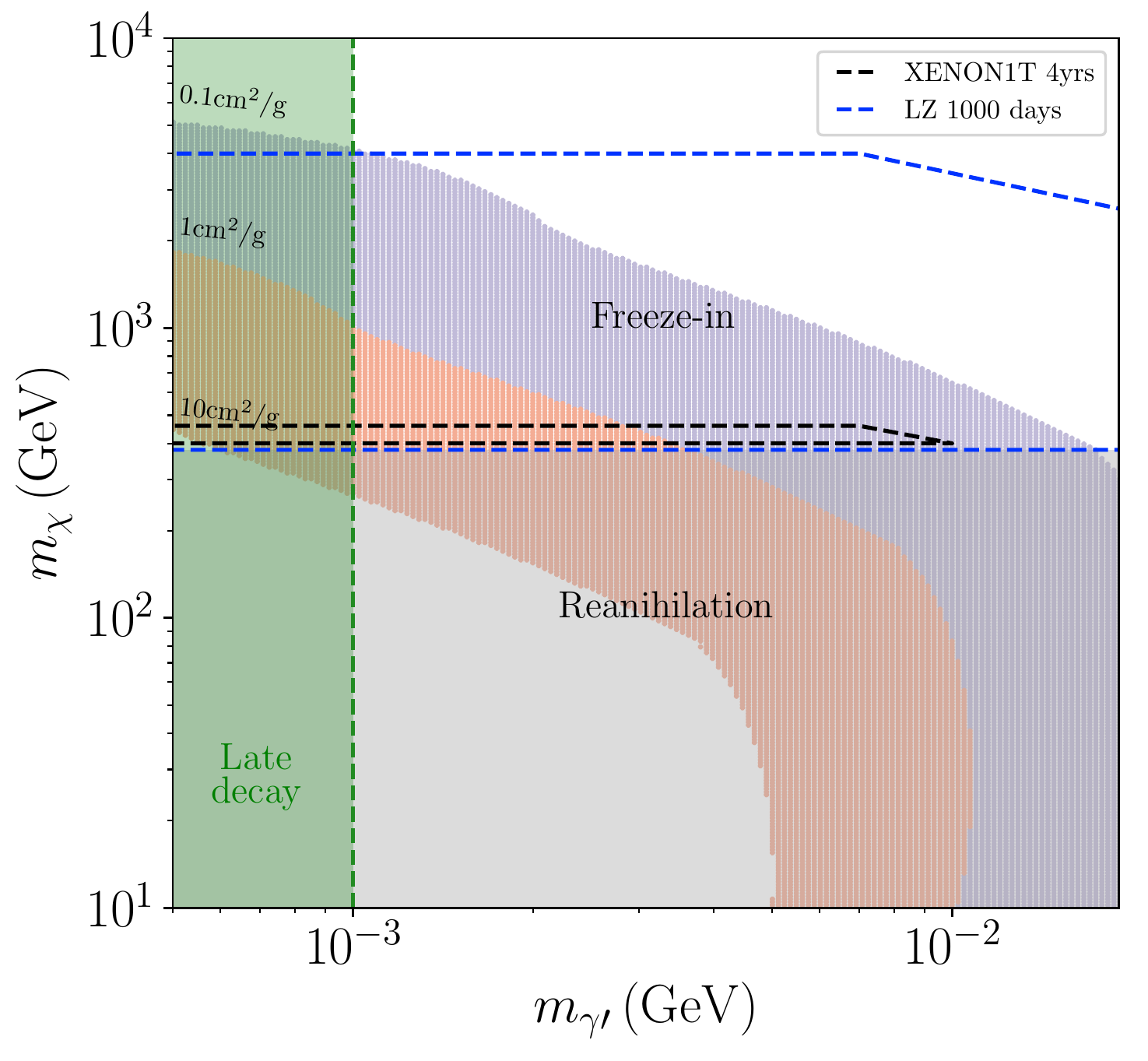}
    \label{f2}
\end{minipage}
\caption{Parameter space for fixed DM coupling and $\alpha^\prime=10^{-3}$ showing DM candidates with attractive (top) and repulsive (bottom) self-interaction that can alleviate the small-scale structure issues. Red: $1 \text{ cm}^{2}/\text{g}\leq\sigma _T/m_\chi \leq 10 \text{ cm}^{2}/\text{g}$; Blue: $0.1 \text{ cm}^{2}/\text{g}\leq\sigma _T/m_\chi \leq 1 \text{ cm}^{2}/\text{g}$. The lines delimitate the FIMP region probed by XENON1T after 4 hypothetical years of exposure (black dashed) and the future LZ (blue dashed). The shaded region corresponds to the reannihilation region which is already probed by XENON1T experiment.}
\label{fig:SIDM2}
\end{figure}

For a larger value of the coupling $\alpha^\prime$ and for a non too larger DM mass, the HS thermalizes and the relic abundance cannot be obtained through the FI regime. One is instead in the so-called reannihilation regime \cite{Frigerio:2011in}. This regime corresponds to a situation in which interactions within the hidden sector are fast compared to the expansion rate, while the energy transfer from the visible sector is slow, so that the temperature of the hidden sector is less than that of the visible sector, see \cite{Frigerio:2011in} for more details. We show in Fig.\ref{fig:SIDM2} the candidates that alleviate the small scale structure problem for $\alpha '=10^{-3}$. In this case, the current XENON1T limits are only probing reannihilation cases. However, Xenon1T should start probing the FI regime after 4 hypothetical years of exposure and this in the range between $m_{\chi} \approx 400$ GeV and $m_{\chi} \approx 480$ GeV. LZ experiment will test FI from $m_{\chi} \approx 400$ GeV up to 4 TeV.

We emphasize that  another example of scenario with a light mediator we could have considered to show that direct detection is testing FI, with possible connections to self-interactions, is the one of Ref.\cite{Bernal:2015ova}, based on the Higgs rather than the kinetic mixing portal. The major difference between the Higgs and kinetic mixing  scenarios is that the former are  constrained  
by Big Bang Nucleosynthesis (BBN). Indeed, a light scalar mediator in the $\sim1$-$100$~MeV range is probably  excluded \cite{Bernal:2015ova} because in this case the decay of the light mediator into an electron and a positron is still active during BBN, as a result of the electron Yukawa coupling suppression of this decay. To determine if, for a scalar portal, there exists a region combining FI, self-interactions and direct detection testability  is still to be investigated further. 
Other works are dealing with direct detection  and self-interacting DM, see {\em e.g.} \cite{DelNobile:2015uua} and/or dark photons, see {\em e.g.} \cite{An:2014twa}, but with a distinct emphasis. The latter deals with DM in the form of dark photons. The former rests on the necessity to suppress, through weak decay into SM particles, the cosmic abundance of dark photons to avoid over-closure of the Universe. We stress that our millicharged FI scenario is meant to be illustrative. We could, for instance, have considered a scenario based on Higgs portal, like in \cite{Bernal:2015ova} (see also \cite{Bernal:2017kxu,Evans:2017kti}), instead of kinetic mixing to show that direct detection is testing FI, with possible connections to self-interactions.

\section{Conclusions}
\label{sec:con}

In this work, we have brought together three corners of DM phenonemology: the FI mechanism for DM creation in the Early Universe, direct detection of DM from the galactic halo and self-interacting DM as a way to address the small scale structures of collisionless DM. In  particular, we have shown that XENON1T data, based on 1 tonne$\times$ year exposure, is testing for the first time the parameter space of a very simple scenario of FI,  based on  millicharged DM coupled to the SM through a dark photon with {mass below $\sim 40$~MeV}. The very same dark photon may induce very large self-interaction for DM, in a range $0.1\,\text{cm}^2/\text{g}<\sigma_T/m_{\rm DM} <10\,\text{cm}^2/\text{g}$. Such FI scenario has several interesting features, and is currently less constrained than the case of FO. Similar results could be anticipated also for other light mediator FI scenarios, such as based on a Higgs portal.

Our main result can be read from Fig.~\ref{fig:Constraints_plot}  where we confront, in the plane $m_{\chi}-\kappa$, FIMP candidates with an abundance that matches cosmological observations, to the exclusion limits from the XENON1T and PANDAX II direct detection experiments, including the reach of LZ (for 1000 days) and XENON1T (for 4 years). These limits have been obtained by recasting the constraints from these experiments on the DM-nucleon SI elastic collision cross section based on a contact interaction, or equivalently mediated by a heavy particle. The approach we have used is straightforward, but is only approximate and does not take full account of the data available to the DM direct detection experimentalists. Given the relevance of the FI mechanism and of its possible relation with self-interacting DM, we believe that it would be important that direction detection collaborations provide in the future their own bounds on scenarios involving a collision cross section proportional to $1/E_R^2$.

\medskip

\begin{acknowledgments}
This work is supported by the FRIA, the "Probing dark matter with neutrinos" ULB-ARC convention, the IISN convention 4.4503.15 and
the Excellence of Science (EoS) convention 30820817.
\end{acknowledgments}

\bibliographystyle{JHEP}

\bibliography{biblio}

\end{document}